\documentclass[preprint,12pt,number]{elsarticle}

\usepackage{subcaption}
\usepackage{hyperref} 
\usepackage{color}
\usepackage{amsmath}
\graphicspath{{Images/}{../Images/}}
\usepackage{caption}

\usepackage{listings}
\definecolor{lightgray}{rgb}{.9,.9,.9}
\definecolor{darkgray}{rgb}{.4,.4,.4}
\definecolor{purple}{rgb}{0.65, 0.12, 0.82}
\lstdefinelanguage{cuda}{
  keywords={const, typeof, new, true, false, catch, function, return, null, catch, switch, var, if, in, while, do, else, case, break, for, call},
  keywordstyle=\color{blue}\bfseries,
  keywords=[2]{boolean, string, number, objectid},
  keywordstyle=[2]\color{green}\bfseries,
  identifierstyle=\color{black},
  sensitive=false,
  comment=[l]{//},
  morecomment=[s]{/*}{*/},
  commentstyle=\color{purple}\ttfamily,
  stringstyle=\color{red}\ttfamily,
  morestring=[b]',
  morestring=[b]"
}
\lstMakeShortInline[columns=fixed]|

\usepackage{amssymb}

\journal{Journal of Parallel and Distributed Computing}

\newcommand\edit[1]{{{#1}}}

\begin{document}

\begin{frontmatter}


\title{Two-Dimensional Batch Linear Programming on the GPU}

\author{John Charlton\fnref{cor1}}
\ead[url]{http://staffwww.dcs.shef.ac.uk/people/J.Charlton/}
\ead{j.a.charlton@sheffield.ac.uk}
\author{Steve Maddock}
\ead{s.maddock@sheffield.ac.uk}
\ead[url]{http://staffwww.dcs.shef.ac.uk/people/S.Maddock/}
\author{Paul Richmond}
\ead{p.richmond@sheffield.ac.uk}
\ead[url]{http://paulrichmond.shef.ac.uk/}
\fntext[cor1]{Corresponding Author}

\address{Department of Computer Science, University of Sheffield, Sheffield S1 4DP, UK. \copyright 2019 This manuscript version is made available under the CC-BY-NC-ND 4.0 license http://creativecommons.org/licenses/by-nc-nd/4.0/ }

\begin{abstract}
This paper presents a novel, high-performance, graphical processing unit-based algorithm for efficiently solving two-dimensional linear programs in batches. The domain of two-dimensional linear programs is particularly useful due to the prevalence of relevant geometric problems. Batch linear programming refers to solving numerous different linear programs within one operation. By solving many linear programs simultaneously and distributing workload evenly across threads, graphical processing unit utilization can be maximized. Speedups of over 22 times and 63 times are obtained against state-of-the-art graphics processing unit and CPU linear program solvers, respectively. 

\end{abstract}


\end{frontmatter}


\section{Introduction}

Linear programming is the challenge of finding an optimal solution to a linear function subject to linear constraints. It is used in many eclectic areas, such as in business problems to maximize profit and in optimal job scheduling. A specialization of this is low dimensional linear programs (LPs) in two or three dimensions. These often have spatial constraints and are used to solve spatial problems such as collision detection between geometric shapes, collision avoidance in pedestrian simulations, and creating cast shapes for moulds. If the speed of solving LPs can be increased, larger problems could be solved faster and real-time constraints could be satisfied in more complex visual simulations, e.g. for fast accurate collision detection.

Graphics processing units (GPUs) are increasingly being used to accelerate performance of data and computationally parallel tasks by offloading such tasks from the CPU. GPUs are a potential candidate solution for improving performance of LP solving due to the inherent levels of parallelism. In general, the theoretical peak performance of GPUs far outperforms similar generation CPUs. As such, data parallel implementations on GPUs can produce much faster code. However, achieving optimal performance of GPU code requires understanding of the architecture to maximize data parallelism, transfer bandwidth and computational parallelism, while hiding latency \cite{brodtkorb_graphics_2013}. 

The general approach to GPU implementations for solving LPs is that data and constraints are initialized on the CPU, transferred to the GPU for calculation and transferred back to the CPU for further processing (or kept on device for other data-intensive tasks and visualisation). A downside to using the GPU in this way is that extra time must be taken to transfer data between devices. Nonetheless, for dense problems of sizes larger than around 700 constraints and 700 dimensions, the computational benefits of using the GPU for solving single LPs mean that, for dense problems, GPU methods outperform CPU methods \cite{bieling_efficient_2010}. Below this amount, or for more sparse problems, any speedup obtained by performing computation on the GPU is offset by memory transfer and associated overhead. A solution to this is to solve large numbers of smaller-sized LPs simultaneously. This increases device utilization and exposes greater parallelism. Simultaneous LP solving, or batch solving, ensures the cost of data transfer to the device is amortised by the gain in computational performance improvements. Gurung and Ray \cite{gurung_solving_2016} report a performance increase for batches as small as 5x5 (5 dimensions and 5 constraints).

The focus of this paper is on two-dimensional LPs solved in batched amounts. We present a Randomized GPU batch (RGB) approach based on the randomized incremental LP algorithm \cite{seidel_small-dimensional_1991}. A naive implementation of the algorithm for parallel architectures is shown to contain both computational and memory imbalances. This paper presents algorithmic improvements to ensure high memory bandwidth performance and balanced workload. The approach taken to ensure computational balance is with subdivision and redistribution of work units, which are sections of code with a similar amount of computation and data requirements. These work units are distributed across the device using ideas inspired from big-data graph analytic algorithms (i.e. cooperative thread arrays) \cite{jog_owl_2013, wang_gunrock:_2016}. This balances workloads between threads within a warp, the GPU's SIMD execution unit, i.e. the collection of threads executing instructions in lockstep. The proposed algorithm can efficiently solve varied batch sizes by offloading work units of larger problems onto threads which are computing smaller problems. The code is available online at \url{https://github.com/coolmule0/LP}.

The algorithms proposed within this paper outperform state-of-the-art CPU and GPU implementations for two-dimensional  problems in cases where the problem size is sufficient to fully occupy the GPU device. Problem size can be increased with larger numbers of constraints or through an increase in batch size which represents the number of simultaneous LPs to solve. A typical example of a two-dimensional problem is physical collision response, e.g. in pedestrian crowd simulations. Here, pedestrians may be represented in two-dimensional space as points with a fixed radius size. To avoid collision with all other pedestrians, each person must solve an LP where each constraint is due to a neighbouring pedestrian. This creates a batch of LPs, one for each person being simulated. Once all the LPs are solved, each person has a new velocity to take which avoids collision. Positions are updated with this new velocity, and the calculation is repeated for the next time step. Due to the repeated nature of this calculation, even a small time reduction in solving LPs is amplified by the repeated solving process for each time step of a simulation.

\edit{To evaluate the implementation of the proposed RGB algorithm, we have benchmarked it against a batch linear GPU simplex solver \cite{gurung_solving_2016}, a multicore CPU solver, CPLEX \cite{ilog_ibm_2009}, and two open-source CPU solvers, GLPK \cite{makhorin_glpk_2008} and CLP \cite{coin_coin-or_2011}. The batch linear GPU solver was chosen as the only one of its kind. CPLEX, GLPK and CLP were chosen due to being high performance solvers \cite{mittelmann_benchmark_2018, meindl_analysis_2012}. We show performance increases versus these solvers for a variety of randomly generated two-dimensional constraints and batch amounts.}

The organization of the paper is as follows. Section \ref{sec:background} covers background theory and related work. Section \ref{sec:algorithmology} explains the novel algorithm implementation and use of thread level work distribution to optimize device utilization. Section \ref{sec:results} presents performance comparisons of the algorithm against two state-of-the-art LP solvers and section \ref{sec:discussion} discusses the results in greater detail. Section 6 gives the conclusions.

\section{Background} \label{sec:background}

\subsection{Linear Programming} \label{sec:lptheory}

Linear programming is the problem of maximizing an objective function subject to linear constraints. The objective function to maximize is represented as

\begin{equation} \label{eq:optimization}
\edit{\text{max} \ \boldsymbol{c}^T \boldsymbol{x} }
\end{equation}
 \edit{where $\boldsymbol{x} = (x_1,...,x_n)$ is the set of values to be determined, and $\boldsymbol{c} = (c_1,...,c_n)$ is the objective function coefficients. $n$ is the dimensionality of the problem.} This is subject to the linear constraints

\begin{equation} \label{eq:constraints}
\edit{\boldsymbol{A}\boldsymbol{x} \leq \boldsymbol{b} }
\end{equation}
\edit{where $\boldsymbol{b} = (b_1,...,b_m)$ is a constant for each constraint, the maximum possible value of the equation, $m$ is the number of constraints of the problem, and $\boldsymbol{A}$ is a matrix of known constraints of size $n \times m$. This set of constraint vectors creates a convex polytope if there is a feasible solution. A linear programming problem is infeasible if there exists no solution for $\boldsymbol{x}$ that satisfies all of the constraints.} 

Various algorithms are suitable for large dimension problems, the most common of which is the simplex algorithm. \edit{In order to apply the simplex algorithm, the linear programming problem must be rewritten in its standard form. In standard form, comparisons are replaced by equalities with a slack variable.}

Incremental linear programming \cite{seidel_small-dimensional_1991} is an alternative option that is conceptually simple and preferable with respect to performance for low dimensional problems \cite{nayak_randomized_1996}. Incremental linear programming works by considering each constraint incrementally and calculating the intermediate objective function for each added constraint. It requires each step to have a unique and well-defined solution. To ensure this, up to two additional constraints per dimension are added \edit{, $\boldsymbol{x} \leq M$ and $\boldsymbol{x} \geq -M$.} These ensure a finite solution and $M$ is taken as very large so as not to affect the optimal solution.

Two outcomes can occur to the intermediate optimal solution when incrementally considering a constraint: (1) if the optimal solution is already satisfied by the new constraint, no change occurs to the intermediate optimal solution; (2) if the intermediate optimal solution does not satisfy the new constraint, the optimal solution will exist on a point on the new constraint intersecting a previous constraint. In the case of (2), \edit{the algorithm to find the location of the new optimal solution is a set of $i-1$ $(n-1)$-dimensional linear programs, where $i$ is the current number of incrementally considered constraints, $0 \leq i \leq m$. When considering two-dimensional problems ($n=2)$, a set of one-dimensional LPs must be solved.} It has been previously proven that one-dimensional LPs can be solved in linear time \cite{de_berg_computational_2008} --- the time to find a new intermediate optimal solution is proportional to the total number of vertices/constraints currently considered, $O(i)$. 

\edit{The set of $(i-1)$ 1D LPs can be parameterized by a variable $\boldsymbol{\hat{u}}$, where $\boldsymbol{\hat{u}}$ is parallel to the added constraint $A_{i}x= b_i$, labelled $l$. $\sigma(h,l)$ is the $\boldsymbol{\hat{u}}$-coordinate of the intersection point of line $\boldsymbol{l}$ and the considered constraint $ \boldsymbol{A}_{h}\boldsymbol{x}= b_h$, where $ 0 \leq h \leq i$, i.e. $h$ indexes over the previously considered constraints. If there is no intersection then either the constraint at index $h$ can be ignored or the linear programming problem is infeasible.} For each remaining constraint of the set, two values should be remembered, depending on whether $h$ is bounded to the left or right:

\begin{equation} \label{eq:uleft}
u_{\textnormal{left}} = \max_{h=(1,...,i-1)} \{ \sigma(h,l) : l \cap h \textnormal{ is bounded to the left}\} 
\end{equation}

\begin{equation} \label{eq:uright}
u_{\textnormal{right}} = \min_{h=(1,...,i-1)} \{ \sigma(h,l) : l \cap h \textnormal{ is bounded to the right}\}
\end{equation}
where $u_{\textnormal{left}}$ represents the leftmost valid point on the line, if the line was horizontal, and, similarly, $u_{\textnormal{right}}$ represents the rightmost valid point on the line. The program is infeasible if $u_{\textnormal{left}} \geq u_{\textnormal{right}}$, otherwise the solution is either $u_{\textnormal{left}}$ or $u_{\textnormal{right}}$ depending on the objective function.

\edit{Calculation of a new optimal solution is only required when the next constraint to consider renders the current optimal solution as infeasible}. A worst case input set would require a re-computation of the solution for each constraint. In this set, each constraint renders the previous optimum solution invalid. If the order of consideration of the worst case input set was reversed, it would create a scenario where only the first constraint would require recomputing the solution. As such, the order of consideration of the input set is important. \edit{There is no simple strategy to optimally organize the order of constraint considerations.} To achieve the best \textit{expected} runtime, the order of consideration should be selected \textit{randomly}. For a single LP calculated serially the expected run time is $O(m)$. 

\subsection{GPU Parallelism}

GPUs are built as high-throughput devices which are designed to maximize the throughput of the pipeline rather than minimize the latency of individual operations \cite{garland_parallel_2008}, as is the case with CPUs. GPUs achieve this through switching groups of executable units on demand when the appropriate resources are available. For example, when a group of executable units is stalled due to a memory request, another group of threads are context-switched so that computation can be performed. Context-switching hides the memory latency and enables high memory bandwidth. To achieve good utilization of the GPU device, two factors are required to hide the latency: compute utilization (high arithmetic intensity) and memory access patterns (reduce the number of total memory movements through the pipeline).

GPUs programs require many threads which execute the same set of instructions (a kernel) on different regions of data (data parallelism). At the execution level, threads are grouped into batches of 32 threads, known as a warp. If some threads within the warp do not require the computation, they are masked out. The worst case scenario for GPU performance is when only one thread requires a certain computation, leaving the remaining threads masked and effectively idle. Any branching paths of execution within a warp causes all threads to execute all paths, with the appropriate threads stalled (or masked out from performing instructions). In general, any aspect of divergent computation should be minimized to improve compute utilization and maximize performance.

Optimizing memory access can be understood through the use of memory fetches. Memory is fetched through 32 byte level two (L2) cache lines. If a byte of data is required, the corresponding 32 byte cache line where the data resides is transferred from memory. It is therefore important to utilize as much data from as few cache lines as possible. In the worst case scenario, each thread within a warp issues a transfer request to move a unique cache line (i.e. a scattered read). The transfer of many cache lines requires greater bandwidth and increases the latency of total memory movement, reducing the performance of the code.

\subsection{Related Work}

Literature on the use of GPU LP solvers can be separated into \edit{three main topics: early solvers before GPU computational APIs, LP solvers aimed at solving single LPs efficiently, and LP solvers specialized in solving multiple problems simultaneously}. 

Research into the use of GPUs for improving linear programming performance began with the start of GPUs as programming units. Early examples (such as \cite{greef_revised_2005, jung_cholesky_2006, jung_implementing_2008}) showed limited performance improvements over serial algorithms for problems larger than 800 dimensions by 800 constraints. Such early models struggled with limited device memory. With hardware advances larger problems can now be tackled and larger speed-ups can be obtained. The advent of dedicated GPU computation programming languages, such as NVIDIA CUDA \cite{nvidia_programming_2010}, has made it easier to develop efficient LP models on the GPU. 

A large proportion of relevant papers examine high-dimensional problems using simplex algorithms, undoubtedly due to the popularity and efficiency of the model. Hall \cite{hall_towards_2010} provides an overview of early multicore simplex algorithms, concluding that speed-ups can be obtained by parallelising for various problem types, excluding large sparse LPs. The trend of applying the simplex algorithms to GPUs has continued, with speed-ups occurring for many different simplex algorithms including the regular simplex algorithm \cite{lalami_efficient_2011, kipfer_lcp_2007} and the revised simplex \cite{bieling_efficient_2010, spampinato_linear_2009}. Ploskas and Samaras \cite{ploskas_efficient_2015} showed that the Primal-Dual exterior point simplex algorithm is more efficient than the more standard Revised Simplex algorithm on GPU hardware. They achieved this through minimizing CPU-GPU memory transfer, since memory transfer dominates the runtime of such large problems.  They achieved speed-up \edit{for all problems tested compared to the simplex-GPU algorithm}. 

The choice of pivot rule plays an important part in simplex algorithms. Choosing a poor rule can slow down performance and lead to no optimal solution being found. Ploskas and Samaras tested the effects of pivot rules on GPU hardware \cite{ploskas_gpu_2014} and found that GPU versions perform better than the CPU equivalents for problems larger than 500 dimensions  and constraints, for dense constraints. Lalami et al. \cite{lalami_efficient_2011} deals with efficient memory transfer through page-locked host memory which provides higher memory bandwidth. In this case, the overhead of data transfer to the device is hidden through asynchronous transfer and computation by staging the problem into smaller units of work. They demonstrate a speed-up of around 12 times for problems larger than 2000 dimensions by 2000 constraints for the regular simplex algorithm, and a speed-up of 2.6 times for problems of 500 dimensions by 500 constraints. With respect to GPU implementations of non-simplex algorithms, an early implementation by Smith et al. \cite{smith_gpu_2012} demonstrates that the matrix-free interior point algorithm shows some performance speed-up for large sparse matrices. For problems larger than around 16,000 dimensions by 16,000 constraints, the GPU implementation outperformed the CPU multicore equivalent. This was due to the efficiency of computational operations required.

An observation of the previous literature suggests there exists a small size limit at which LPs should only be computed on the CPU. This is due to the limited parallelism available for smaller problem sizes. Below this amount CPU implementations are equivalent or better performing than GPU equivalents. \edit{CPLEX is a CPU optimization software package containing the functionality to solve LPs \cite{cplex_12.2_2010}. It is able to solve problems using different algorithms including dual simplex, primal simplex and barrier method. It uses multithreading to solve models, but this efficiency decreases up to 4-8 threads after which increasing thread count will not significantly change execution time \cite{noauthor_ibm_2013}. GLPK \cite{makhorin_glpk_2008} and CLP \cite{coin_coin-or_2011} are open-source serial simplex solver methods. The performance of different CPU LP solvers is extensively tested \cite{mittelmann_benchmark_2018, gearhart_comparison_nodate, noauthor_ibm_nodate} but can still remain challenging to find the most efficient method for a given problem.}

In order to expose greater parallelism for small LPs,  many small LPs should be computed simultaneously. Gurung and Ray \cite{gurung_solving_2016} examines an algorithm for solving numerous dense LPs simultaneously using the simplex algorithm on the GPU. By considering many LPs at once, asynchronous memory transfer can occur simultaneously with computation of results to increase performance. Also the use of many LPs ensures the computational cores on the device are all being utilized. They show a speed-up over equivalent CPU algorithms for square LPs as small as 5 dimensions by 5 constraints for 100 batches. \edit{Multiple concurrent streams are active in the algorithm. Such streams allow overlapping memory transfer and kernel computation. Thus, rather than copying all the data across from CPU to GPU, then solving, then copying data back to CPU, the batch LPs are split into smaller groupings. Once data has been transferred to the GPU for one of these smaller groups, the kernel can perform the required computation. Simultaneously, the information for another group can be copied across to the GPU. By splitting tasks into these smaller groups, total device utilization is increased, increasing performance. The current implementation limits the size of feasible LPs to 511 dimensions by 511 constraints \cite{gurung_simultaneous_2018}.  }

\section{The Algorithm} \label{sec:algorithmology}

This paper presents a randomized GPU batch (RGB) algorithm \edit{in two-dimensions} that is based on Seidel's incremental linear programming algorithm \edit{\cite{seidel_small-dimensional_1991}, explained in section \ref{sec:lptheory}}. Alterations to Seidel's work were made to improve performance when implemented on a GPU architecture, increasing compute and memory parallelism. \edit{This was achieved by decomposing the calculation into small work units (WU), which are shared between threads in a block to more evenly spread compute and memory load between threads on the device. The algorithm is designed and tested for solving 2D problems. }

\begin{figure}[htb]
\center
    \includegraphics[height=6cm]{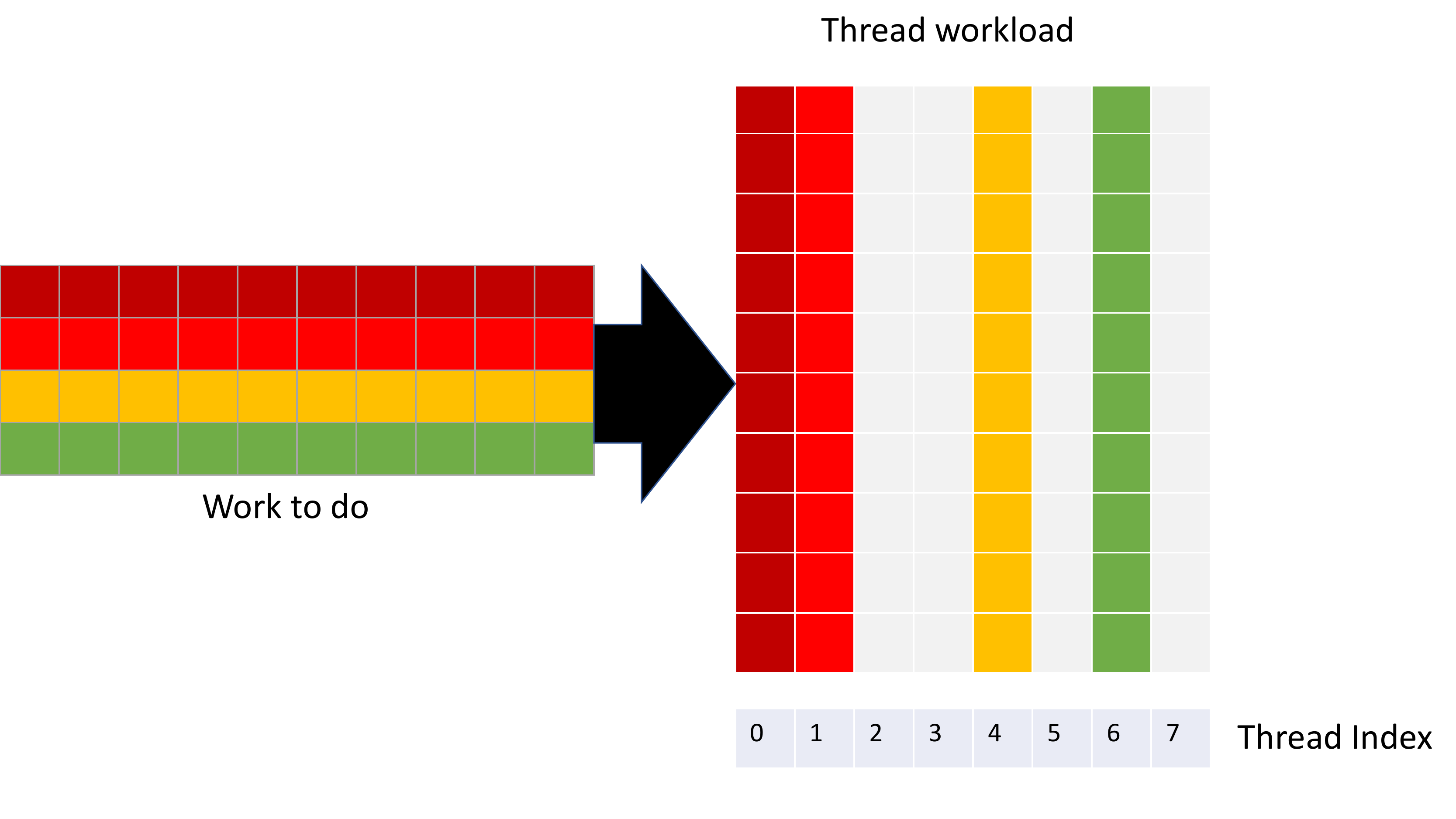}
    \caption{Distribution of workload across a thread before optimizations. Imbalanced workloads reduces the computation parallelism}
    \label{fig:NaiveWork}
\end{figure}

To implement incremental linear batch LP solving on the GPU, each LP is assigned to a core/thread. Thus the number of active threads is equal to the number of LPs to be solved. A naive implementation of Seidel's algorithm would result in a large divergence in calculations between threads, as illustrated in Figure \ref{fig:NaiveWork}. In this figure some threads require large amounts of computation while others require very little. This imbalance can be attributed to all threads within a warp considering the same index of constraint. In this context, some problems (threads) will be satisfied with the considered constraint while others need to recompute the intermediate optimal solution. This divergence causes an imbalance of workload within a warp and hence results in poor performance.

To address the problem of imbalance, the idea of cooperative thread arrays can be applied \cite{jog_owl_2013, wang_gunrock:_2016}. In this case, threads within a warp or block communicate with each other to share the workload. In the RGB algorithm the most intensive computational aspect is performing the set of 1D linear programs for all previous constraints -- see equations (\ref{eq:uleft}) and (\ref{eq:uright}). This was found through timing, profiling and analyzing of code performance. Each constraint to perform a 1D LP can be thought of as the smallest quantity of work, referred to as a work unit (WU). These WUs can be distributed across threads in a block so parallel computation is more balanced. The requirement for this is that the writing of the results $u_{\textnormal{left}}$ and $u_{\textnormal{right}}$ must be done atomically into shared memory. Atomic writing ensures no race conditions occur for the results. This ensures the correct result but reduces performance compared to a standard write to memory. Other techniques are available to use, such as reduction, but atomic operations work well for unknown set sizes at runtime, and atomic operations have improved in performance with recent hardware \cite{nvidia_tuning_2018}. \edit{Examination into shared memory atomics shows that on Maxwell hardware shared memory atomics outperforms global atomics \cite{sakharnykh_gpu_2015} and device-wide segmented reduction. Shared memory atomics also has stable performance across a range of workloads, an important aspect for the RGB algorithm where many different amounts of contention are present. See Figure \ref{fig:atomicComp} in Section \ref{sec:results} for computational results. }

Communicating data between threads in a block is done through shared memory, a region of memory that any threads in the same block can access on chip. Threads will read appropriate constraint data from global memory, stored on device DRAM, and write it to shared memory. The data is stored in shared memory due to low latency access within a block. Since different threads make numerous reads to the data, shared memory is far more efficient than the alternative global memory, which takes around 100 times longer to access \cite{nvidia_programming_2010}. Shared memory is limited in size so only the most accessed pieces of data are loaded in. The remaining data is loaded from global memory and access to such memory is coalesced as much as possible to avoid excessive cache line fetches. Vectorized loads are used to reduce the number of memory requests and increase the utilization of cache lines where any scattered reads are required. Since information of half-planes is stored in multiple variables (2D position and direction) combining the information into one extended set of data ensures that scattered reads uses as much of each cache line as possible. Memory transfer between CPU and GPU is managed through the use of CUDA Managed Memory. This lets the underlying runtime handle the paged transfer of data to and from the device intelligently. It removes the requirement that all data be copied to device at kernel launch, instead paging in memory as demanded and asynchronously during kernel execution. This reduces the time of copying memory to and from the device. It also allows for allocating memory up to the size of system memory, rather than being restricted to a maximum of the dedicated GPU DRAM. This is important in large problems when the device DRAM is too small, but hardware DRAM is large enough.

\begin{figure}[htb]
\center
    \includegraphics[height=6cm]{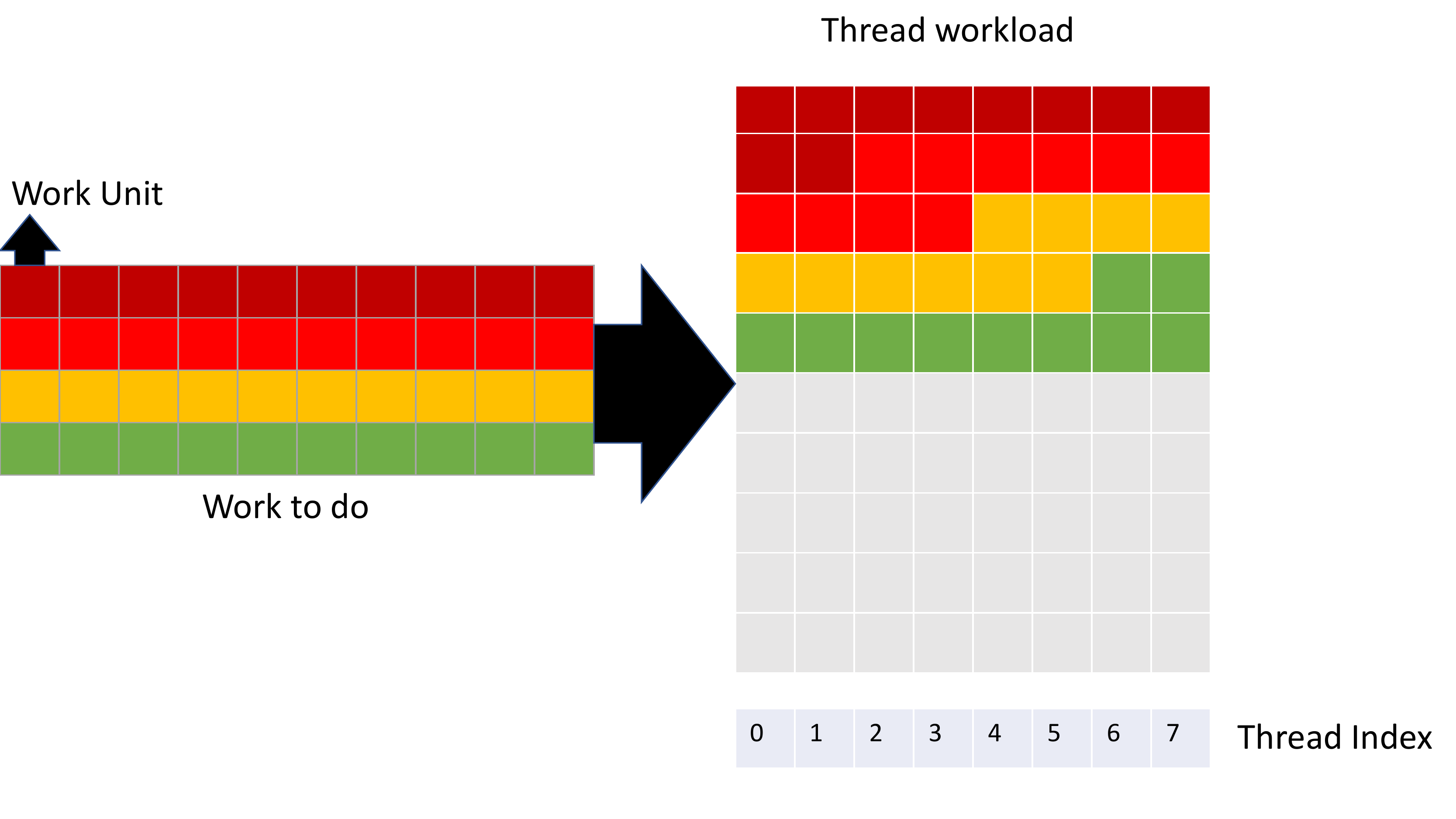}
    \caption{Distribution of workload across a warp after optimizations. Work units are distributed evenly across all available threads.}
    \label{fig:NewWork}
\end{figure}

Figure \ref{fig:NewWork} highlights the use of cooperative thread arrays. The amount of work required is computed and distributed across all available threads. The sharing of data is done through shared memory. This creates more even workloads which increases the parallelism of the problem and reduces the overall time for computation.

An overview of the algorithm is provided in Listing \ref{code:c-code}. The algorithm runs as many threads as there are problems to solve, |p|. The program incrementally examines each constraint in a problem and loops over the maximum problem size. The current line to consider is read into shared memory from global device memory. This read uses coalesced memory access for optimal performance. It checks whether the LP solution, stored in |S|, is satisfied by this line in |unsatisfied| and writes this logical check to a binary value |B|. A |sync_block()| call ensures all data has been written to before it is accessed, avoiding race conditions. The binary value |B| is reduced to calculate the total number of problems in need of recomputation in the block. All threads are mapped to a work unit, a unique tuple of problem and line, and this is repeated over all lines for each problem in need of recomputation. The work unit calculation reads in the assigned line from device memory, |line|, and calculates the intersection point between this |line| and the considered line in shared memory, |SML|. The result is written to shared memory using atomics to avoid race conditions. Another |sync_block()| call ensures all calculations are complete before updating the solution.

\begin{lstlisting}[escapeinside={|}{|},caption={RGB GPU Algorithm Overview},label={code:c-code}]
//p is batch problem size
//lp_max is maximum LP size
//L is a device vector of m |$\times$| lp_max line constraints containing all constraint lines
//SML is an empty shared memory vector of length m, containing current constraint line
//S is an empty device vector of length m, containing the solution
//SMS is an empty shared memory scalar of length threads in block,
//B is an empty shared memory scalar of length threads in block
//block_width is the width of the CUDA Kernel block configuration
gpu_parallel_for idx |$\leftarrow$| 1 to m 

    bidx |$\leftarrow$| idx |\%| block_width

    for n |$\leftarrow$| 1 to lp_max
        SML[bidx] |$\leftarrow$| L[idx][n]
        B[bidx] |$\leftarrow$| call unsatisfied(SML[bidx], S[idx])
        call sync_block()
    
        active_threads |$\leftarrow$| block_reduce_sum(B)
        wu_count = active_threads * n
        
        j |$\leftarrow$| bidx
        while j < wu_count
            map_idx, map_n |$\leftarrow$| map(j)
            map_bidx |$\leftarrow$| map_idx |\%| block_width
            line |$\leftarrow$| L[map_idx][map_n]
            i_pos |$\leftarrow$| call intersect(SML[map_bidx], line)
            SMS[map_bidx] |$\leftarrow$| call atomicMin(SMS[map_bidx], i_pos)
            j |$\leftarrow$| j + block_width
        
        call sync_block()
        
        if not B[bidx]
            S[idx] |$\leftarrow$| SMS[bidx]
\end{lstlisting}

%
%
%
%
%
%
%
%
%

\section{Results} \label{sec:results}

\edit{In this section, results obtained from running the RGB algorithm are presented and compared against four other algorithms:
(i) an open-source CPU algorithm, the GNU Linear Programming Kit (GLPK) \cite{makhorin_glpk_2008}; (ii) CLP, a high-performing \cite{mittelmann_benchmark_2018} open-source CPU simplex solver \cite{coin_coin-or_2011}; (iii)  CPLEX, a high-performing \cite{mittelmann_benchmark_2018} multi-core CPU solver \cite{ilog_ibm_2009}; (iv) the batch-GPU simplex algorithm of Gurung and Ray \cite{gurung_solving_2016}, used as an algorithm aimed at solving batched LPs on the GPU. The naive implementation of the RGB algorithm without improvements (as described in section \ref{sec:algorithmology}) is referred to as `NaiveRGB' and is also tested.} 

\edit{ The GLPK algorithm is parallelized over LPs, allowing different threads to solve separate problems, and is referred to in the results as `mGLPK', standing for `multicore-environment-GLPK'. CPLEX is able to use different methods to solve LPs -- in the tests, the automatic algorithm selector was used, which allows the underlying solver to choose which algorithm it believes is most suitable. CLP is a single-core solver and is set to solve using the dual simplex method. Tests were also run for GLPK in a serial manner. However, the results of this are not shown as performance for the multicore-environment version (mGLPK) was better, with improvement in performance of up to 6 times, i.e. the number of cores on the CPU tested. }

\edit{ Tests were run on an NVIDIA Titan V  GPU card with 12GB dedicated memory and a (6 core) Intel i7-6850K with 64 GB RAM. The GPU was connected by PCI-E 2.0. The GPU software was developed with NVIDIA CUDA 8.0 on Ubuntu, and CPU code was compiled with gcc 7.3. The algorithms were timed after the problem had finished initializing on the CPU, and ended when the result had been written to CPU-usable memory. In the case of GPU timing, this included data transfer to and from device as a result of CUDA managed memory paging. }

Problem sets are generated using random feasible constraints in two-dimensions: constraint lines are generated randomly and tested to ensure a solution is possible. Only one LP is generated per run, and copied multiple times into memory to simulate batch numbers. Due to numerical deviations between CPU and GPU floating point accumulation, a tolerance value of 5 significant figures is set on the results to ensure that consistent results are obtained for all algorithms. \edit{These problems are repeated multiple times with new random feasible problems, with the error bars representing one standard deviation of uncertainty.}

\begin{figure}
\begin{subfigure}{0.5\textwidth}
\center
    \includegraphics[width=0.95\linewidth]{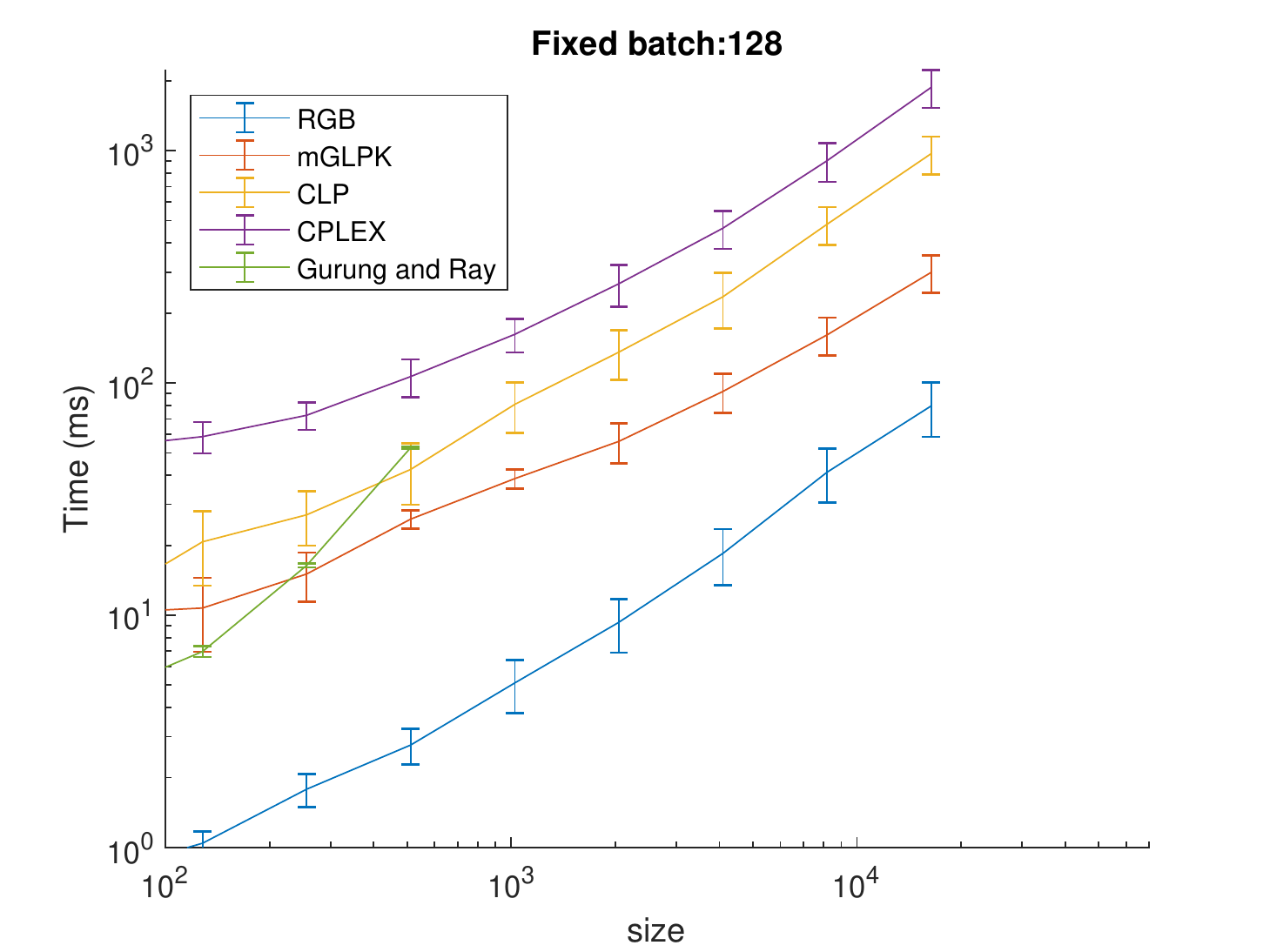}
    \caption{Timing vs varying LP sizes for fixed batch amount 128}
    \label{fig:smol}
\end{subfigure}
\begin{subfigure}{0.5\textwidth}
\center
    \includegraphics[width=0.95\linewidth]{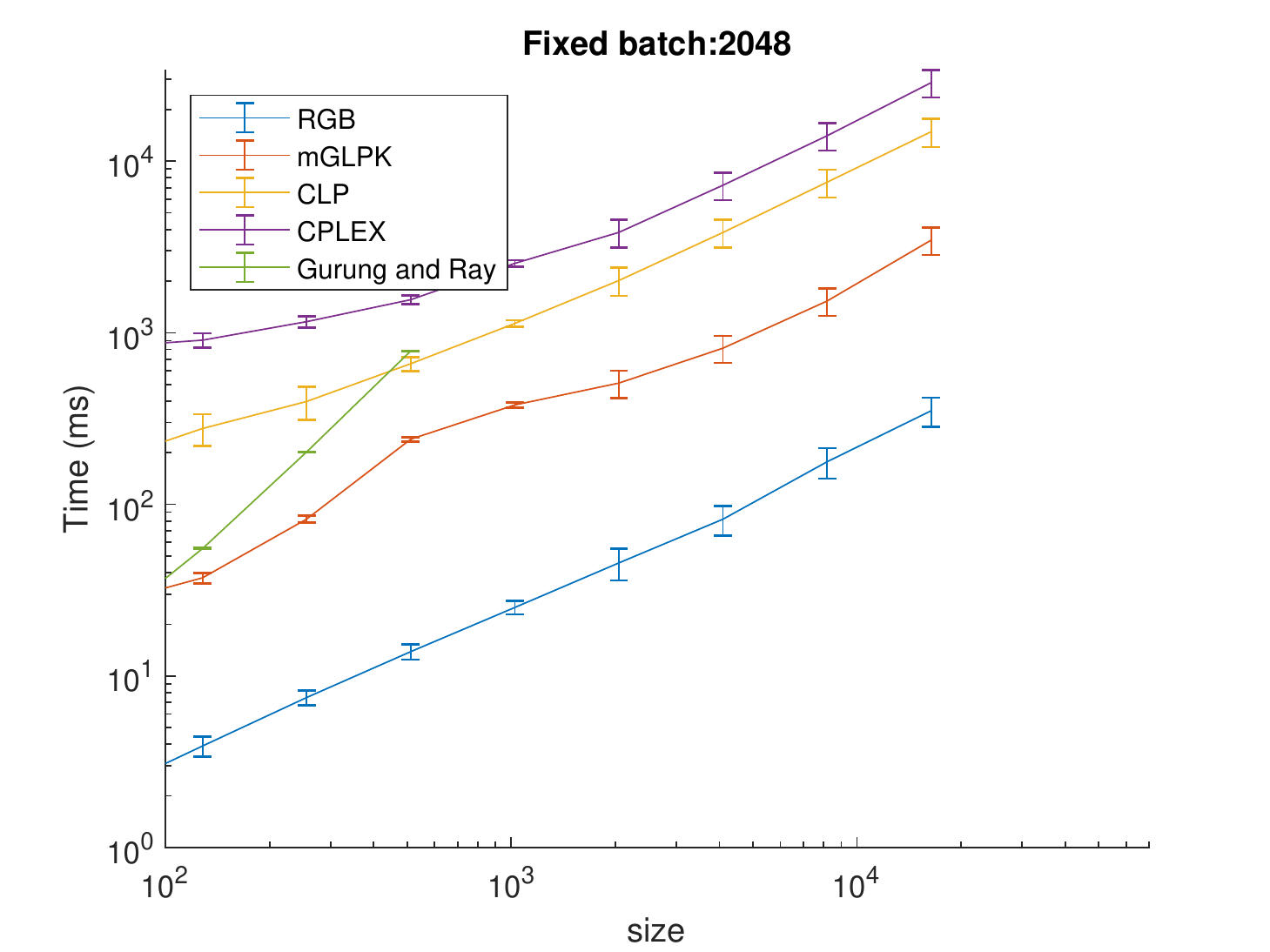}
    \caption{Timing vs varying LP sizes for fixed batch amount 2048}
    \label{fig:med}
\end{subfigure}

\begin{subfigure}{0.5\textwidth}
\center
    \includegraphics[width=0.95\linewidth]{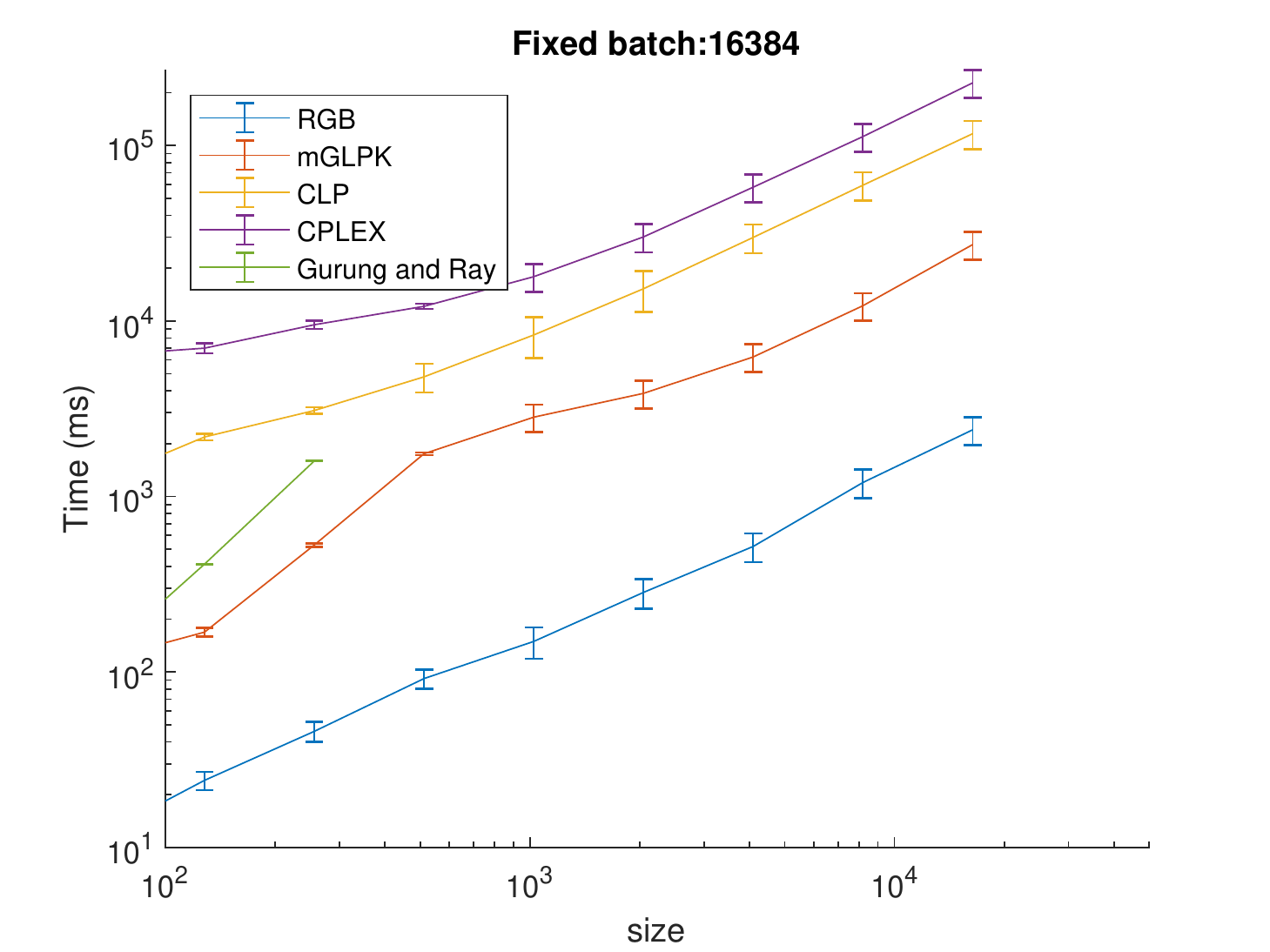}
    \caption{Timing vs varying LP sizes for fixed batch amount 161384}
    \label{fig:larg}
\end{subfigure}
\caption{Timing comparison of the three algorithms for fixed batch sizes and varied LP sizes.}
\label{fig:fixedbatch}
\end{figure}

\edit{The results shown in Figures \ref{fig:smol} - \ref{fig:larg} use fixed batch sizes and measure the time taken to solve all LPs when varying the sizes of problems. The RGB algorithm can be seen to outperform the other algorithms above sizes of $2 \times 10^2$. The algorithm of Gurung and Ray was limited to smaller sizes, and can be seen to improve compared to CPU algorithms as the number of batches increases. }

\begin{figure}
\begin{subfigure}{0.5\textwidth}
\center
    \includegraphics[width=0.95\linewidth]{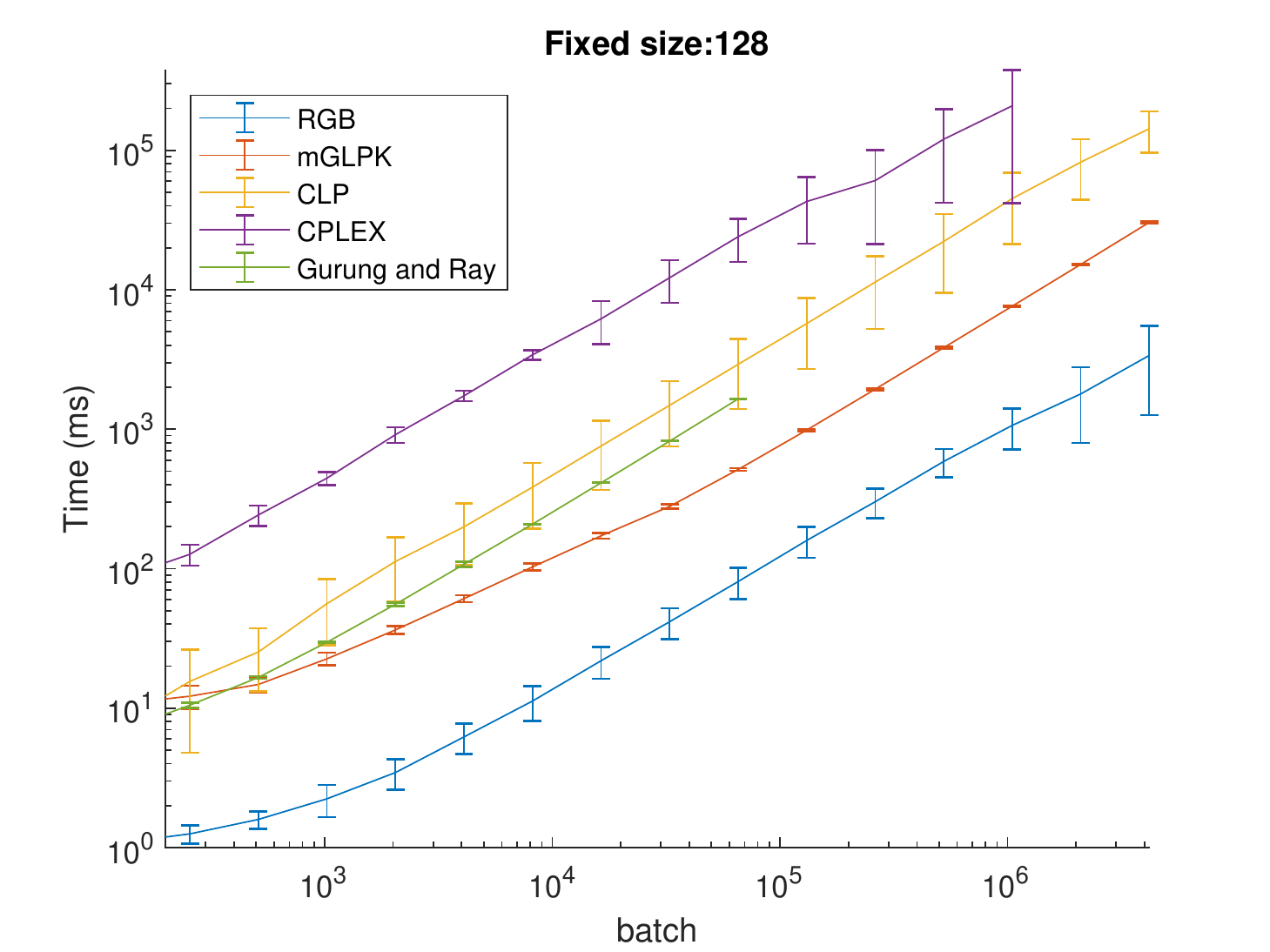}
    \caption{Timing vs varying batch amounts for fixed size 64 constraints}
    \label{fig:smolsize}
\end{subfigure}
\begin{subfigure}{0.5\textwidth}
\center
    \includegraphics[width=0.95\linewidth]{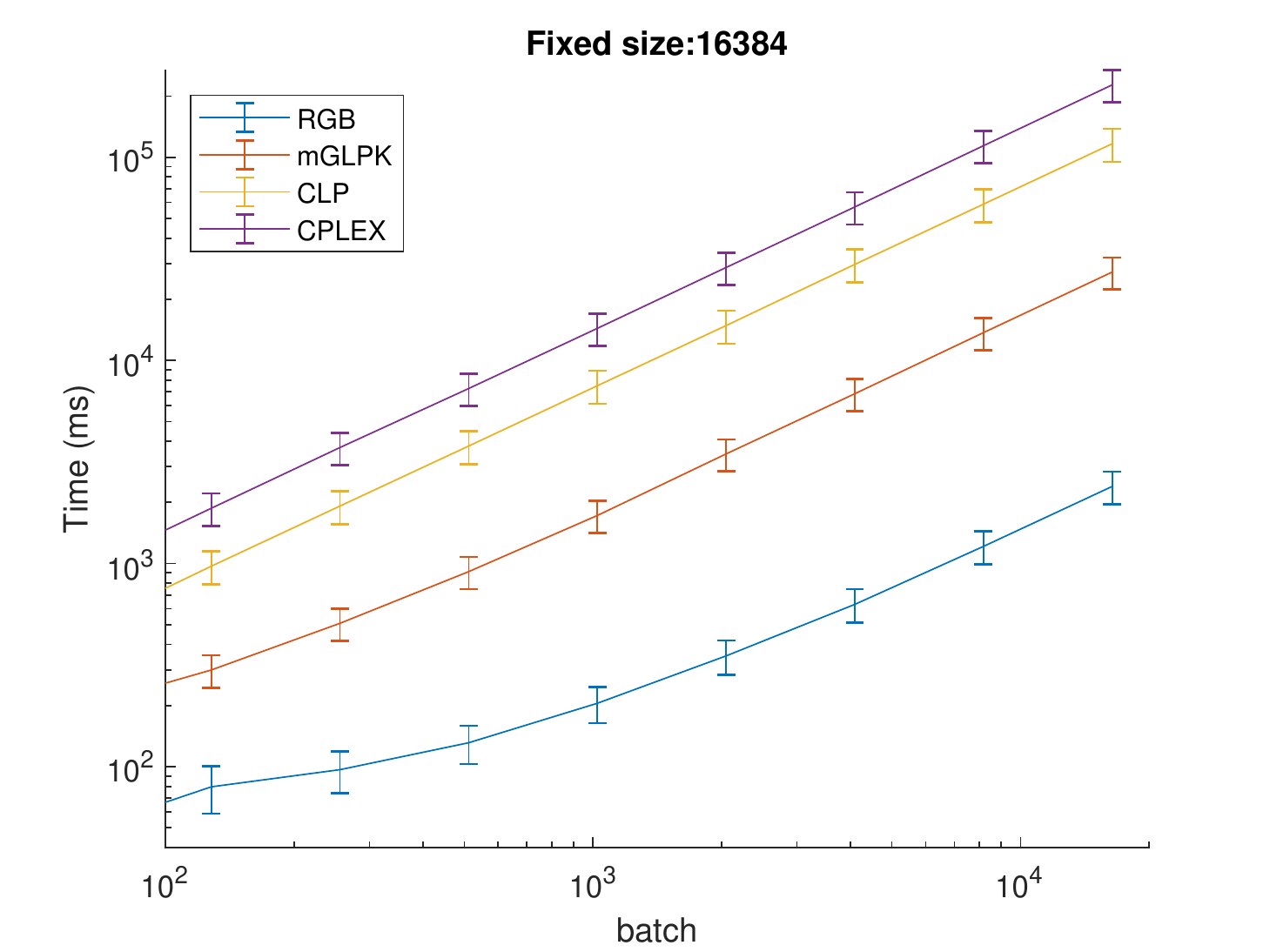}
    \caption{Timing vs varying batch amounts for fixed size 8192 constraints}
    \label{fig:medsize}
\end{subfigure}
\caption{Timing comparison of the three algorithms for fixed constraint size and varied batch amounts.}
\label{fig:fixedsize}
\end{figure}

Figures \ref{fig:smolsize} - \ref{fig:medsize} show results from the same experiment, but with varying batch numbers and fixed problem sizes. In Figure \ref{fig:smolsize}, the RGB algorithm can be seen to outperform the other algorithms for all tested batch sizes with LP sizes of 128. In Figure \ref{fig:medsize}, the RGB algorithm outperforms the CPU implementations for all tested batch sizes. The Gurung and Ray algorithm was not able to be tested at this large problem size.

\begin{figure}
\center
    \includegraphics[height=6cm]{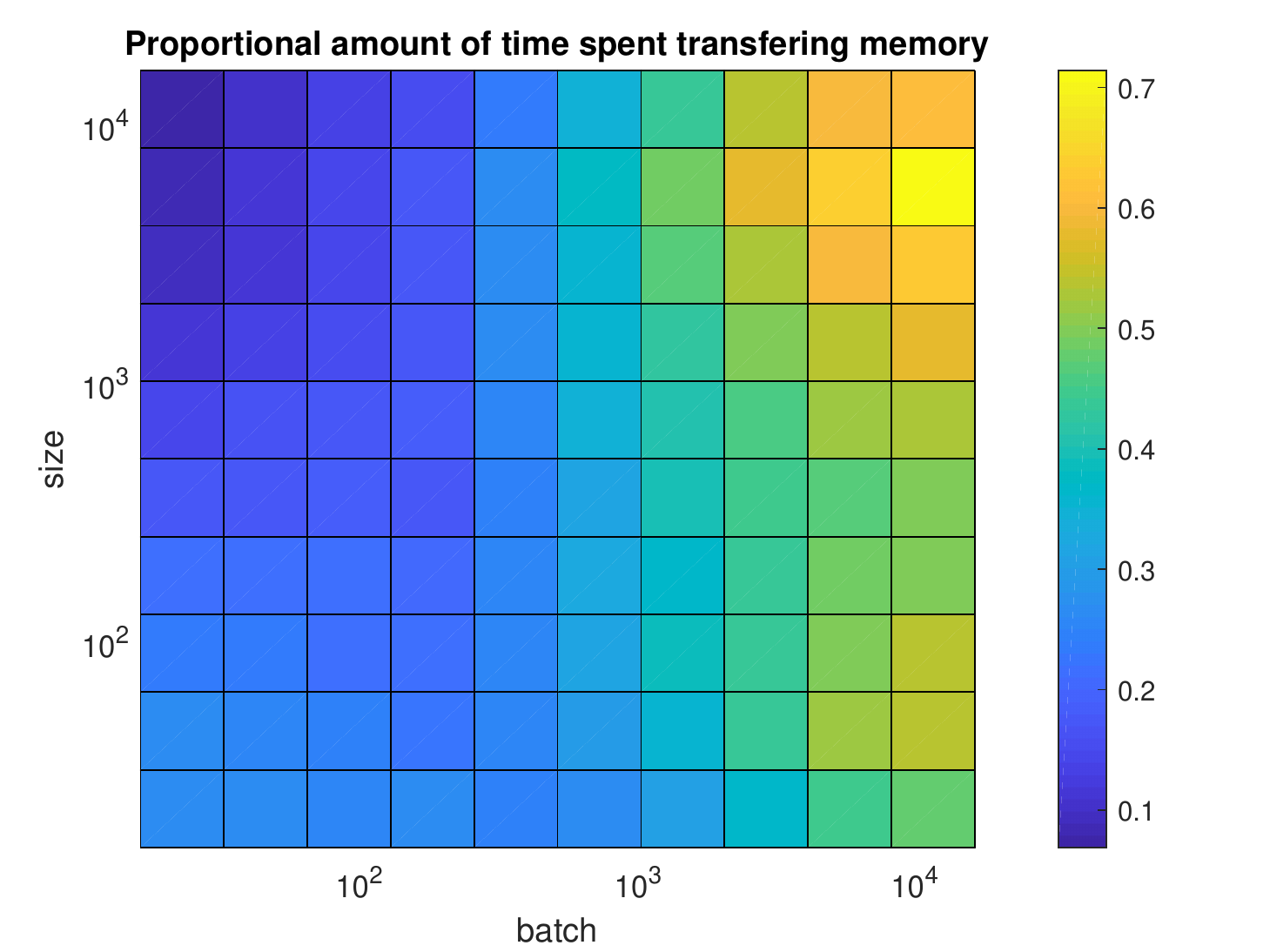}
    \caption{Proportion of time spent copying memory compared to total execution time for RGB}
    \label{fig:memsurf}
\end{figure}

\edit{For large batch sizes the majority of the execution time of the RGB algorithm is due to memory initialization and transfer. For these larger sizes the computation kernel takes less than  $30\%$ of the total execution time, with the remainder being used to manage memory. This is highlighted in the surface plot in Figure \ref{fig:memsurf}. The bright yellow area represents size-batch problems which use more than $40\%$ of the time to initialize and transfer data. The dark blue region is where the majority of time is spent performing computation. It shows that as batch amounts increase, the proportional amount of time spent transferring memory also increases. }

\begin{figure}[htb]
\center
    \includegraphics[height=6cm]{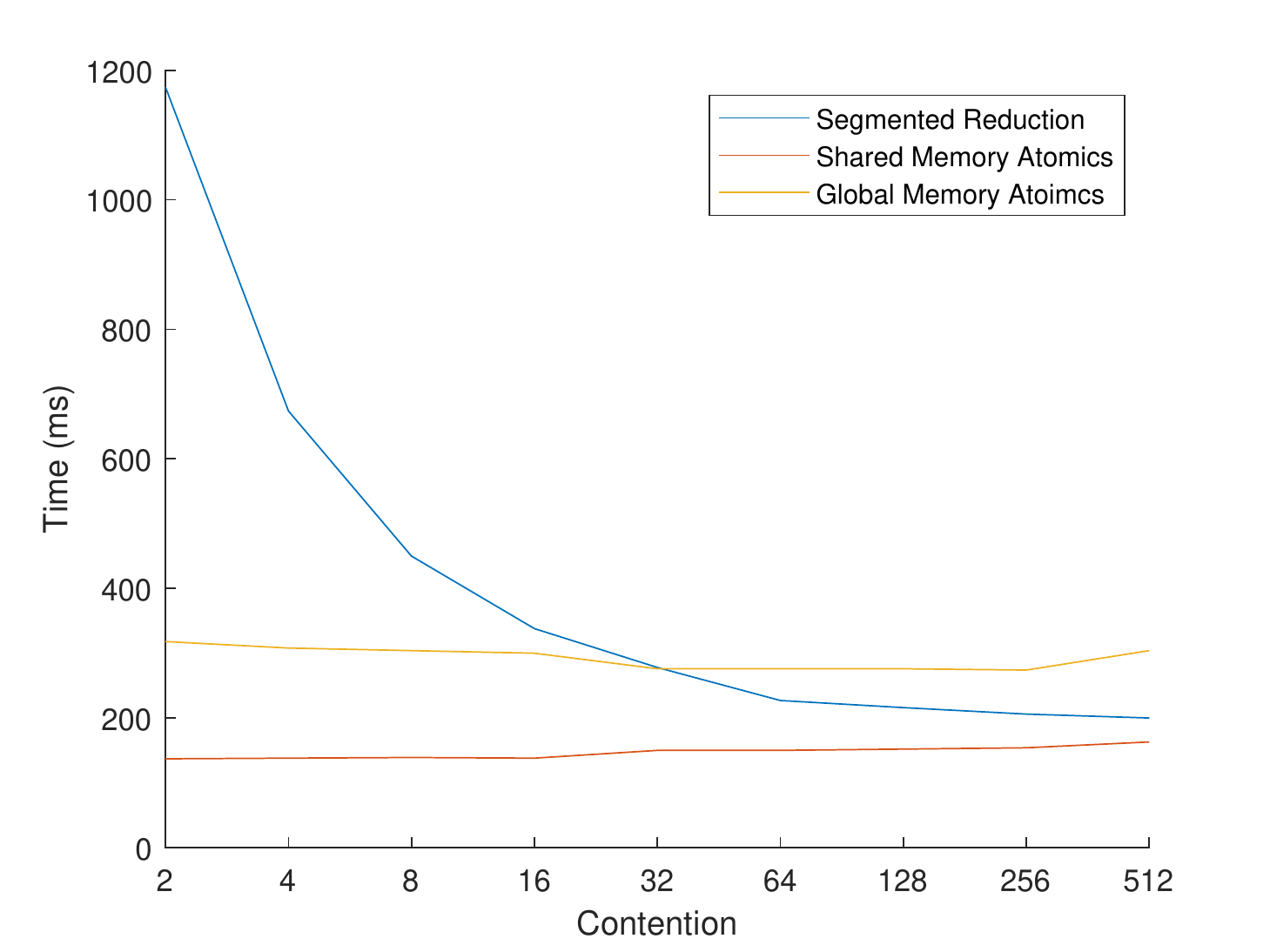}
    \caption{Performance vs contention for atomics and device-wide segmented reduction.}
    \label{fig:atomicComp}
\end{figure}

\edit{An important computational aspect of the algorithm is the performance of the atomic reduction. Figure \ref{fig:atomicComp} shows the performance comparison of shared memory atomics, global memory atomics and device-wide segmented reduction (using the CUB library \cite{merrill_cub_2013}) over a range of contention. Contention is a measure of how many elements must reduce into a final value. A contention of 2 means every 2 elements must reduce their values together. The maximum measured contention is 512, chosen as the block size of the kernel. Results show shared memory atomics to be consistent in timing and better performance in comparison for all contention measured. Shared memory atomics also has stable performance across a range of contention, an important aspect for the RGB algorithm where many different values of contention occur.
}

\begin{figure}
\begin{subfigure}{0.5\textwidth}
\center
    \includegraphics[width=0.95\linewidth]{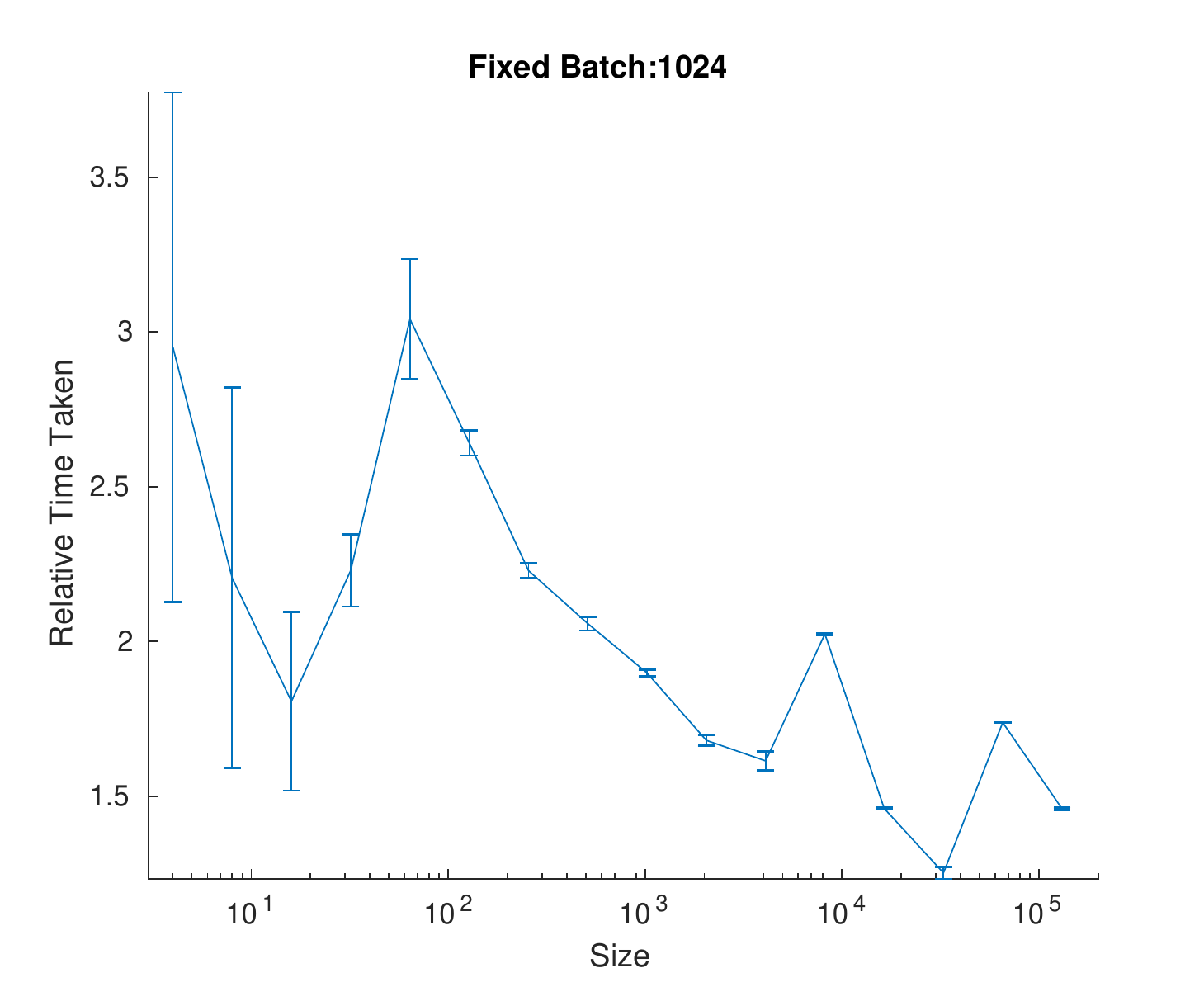}
    \caption{Relative timing of NaiveRGB and RGB implementations vs varying LP sizes for 1024 problems}
    \label{fig:smolrel}
\end{subfigure}
\begin{subfigure}{0.5\textwidth}
\center
    \includegraphics[width=0.95\linewidth]{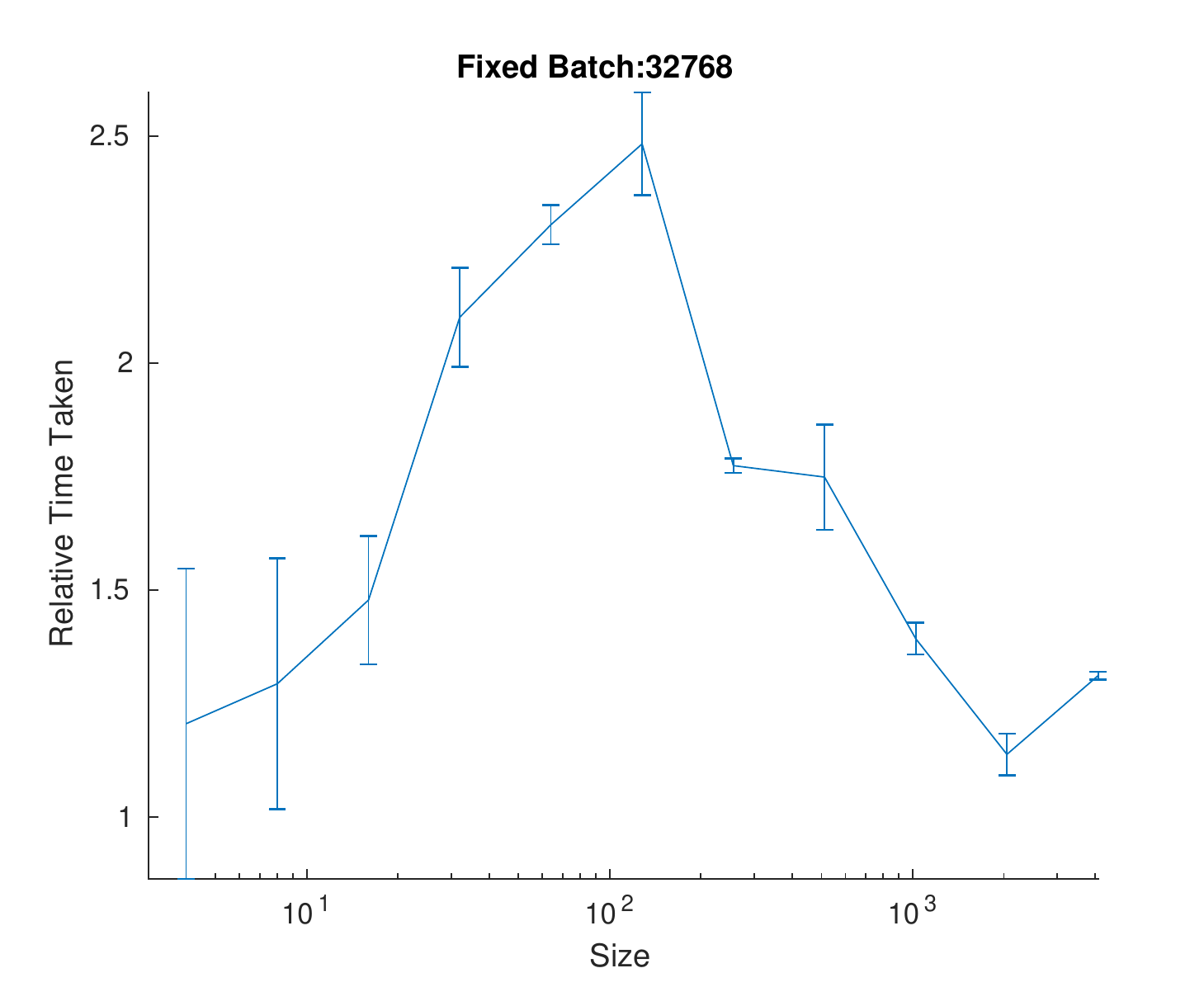}
    \caption{Relative timing of NaiveRGB and RGB implementations vs varying LP sizes for 32768 problems}
    \label{fig:medrel}
\end{subfigure}
\caption{Relative timing comparison of naive NaiveRGB and optimized RGB algorithms for fixed constraint size and varied batch amounts.}
\label{fig:fixedsizerel}
\end{figure}

The performance difference between naive NaiveRGB and optimized RGB is examined in Figure \ref{fig:fixedsizerel}. To highlight the difference in performance, the relative computation kernel execution time of naive NaiveRGB and optimized RGB are shown, which ignores the time taken for data transfer. The y-axis shows the speedup of the optimized RGB algorithm over the naive NaiveRGB implementation, with a value of 1 meaning both algorithms execute in the same time.

\section{Discussion} \label{sec:discussion}

The results show that there is a trend for a greater speed-up to occur for RGB against CPU algorithms as batch amounts increase and LP sizes decrease.  There is a similar trend for a speed-up for RGB against the algorithm of Gurung and Ray, as batch amounts increase and LP sizes also increase. This means that the CPU models scale better to larger LP sizes but does not scale with batch amount. This is expected due to the powerful serial performance of the CPU. This can be seen in figures \ref{fig:smolsize} and \ref{fig:medsize} where the time taken to complete execution increases at a greater rate for mGPLK and CLP than RGB when batch sizes are increased. Figures \ref{fig:smol}-\ref{fig:larg} highlight the difference in time scaling for varying batch amounts. The batch method of Gurung and Ray can be seen to scale much worse than RGB in these figures. This difference in scaling is due to the RGB model using methods suited to two-dimensional constraints, such as repeatedly running one dimensional LPs, which is computationally light compared to the operations used in the batch simplex algorithm. \edit{mGLPK tends to be the best performing out of the CPU models tested due to the multicore environment in which it is run allowing it to be better suited for solving batch amounts.} Maximum speed-ups are reported as 66x for the RGB algorithm against mGLPK algorithm, and 22x for the RGB algorithm against Gurung and Ray's algorithm. The relative performance of the naive and optimized RGB algorithms peaks in relative difference around $10^2$ for figures \ref{fig:smolsize} - \ref{fig:medrel}. This is due to launching block sizes in such a way as to be optimized for this size. Various operations used, such as block compression, requires a fixed block size at compile time. Further improvements to performance of the optimized RGB algorithm can be made by tailoring block sizes to the expected LP size. A limit in improving the optimized RGB algorithm performance is the memory bandwidth limit. Using GPUs with greater memory bandwidth should show increased performance over the other tested methods.

The RGB algorithm is the best performing out of the algorithms tested when the GPU device is being fully utilized. For problems that cannot fully utilize the device's cores, CPU algorithms mGLPK and CLP are more efficient solvers. Full device utilization occurs when there are more constraints to consider than cores in the device, i.e. $\textnormal{No.-of-batches} \times \textnormal{size-of-batches} \gg \textnormal{gpu-cores}$. For the tested hardware, a Titan X Maxwell, this is $3072$. Above this amount all cores are used in the computation. Below full device utilization it is hard to hide the large latency that is inherent on the GPU, which can be hidden by exposing greater levels of parallelism.

A practical use of the RGB algorithm has been applied to an early model of pedestrian simulation. The initial results are able to solve collision avoidance amongst millions of people to provide real time simulation of large-scale crowds. The algorithm was purely GPU based without CPU data transfer, allowing for up to $10^6$ people to be simulated and visualized in real time. Compared to a CPU implementation this early indication suggests performance around 11 times faster for similar numbers of people. Each person represented a constraint for each neighbour, resulting in a batch of LPs, one for each person. These were solved to find an optimal velocity for each person that avoided collisions with all other people and got them to their destination in the fastest time and smoothest path. Additional computation is required due to not guaranteeing LPs to be feasible, e.g. when collision-free motion cannot be guaranteed. This was repeated each time step, so any performance improvements that can be obtained are compounded due to the repeated nature of the simulation. Performance of the algorithm is key to allow for more simulated people while remaining in real time.

\section{Conclusion} \label{sec:conclusion}

This paper has presented a novel implementation of a \edit{low-dimension} batch linear program solver suitable for GPUs. The implementation makes use of cooperative thread arrays to share workload across threads. The RGB algorithm outperforms Gurung and Ray's work \cite{gurung_solving_2016} for all problem sizes above 300 constraints in two-dimensions. It also outperforms the CPU algorithms, GLPK \cite{makhorin_glpk_2008} and CPLEX \cite{cplex_12.2_2010}, for problems that are known to fully utilize the device. 

Timings were measured to include data transfer to and from device (i.e. between CPU and GPU). It should be noted that if the batch-GPU LP algorithm is just one stage of computation that is taking place on the GPU then the time of copying data from CPU to GPU can be improved. This can vastly increase performance at larger data sizes or for iterative problems such as pedestrian simulations and collision avoidance. When utilizing the whole device the problem is memory bound. As such, further improvements to the model should optimize data loading and transference. For this paper the use of CUDA Managed Memory provided sufficient data performance, though this could be improved upon on a per-device basis.

The domain of the model is best suited to solving large numbers of low dimensional LPs, each of which has numerous constraints. An advantage of the model is the allowance for different-sized individual LPs within the batches. \edit{The distribution of work units ensures that workload is balanced regardless of variance in batch problem sizes.} Future directions could examine the applications and performance of the model extended to higher dimensions. \edit{ It is expected to scale favourably for low dimensional problems, up to around 5 dimensions, due to the efficiency of the core solving algorithm at such low dimensions.}

\section{Acknowledgements}
This research was supported by EPSRC grant ``Accelerating Scientific Discovery with Accelerated Computing" (grant number EP/N018869/1) and the Transport Systems Catapult.

\section{References}
\bibliographystyle{ieeetr}
\bibliography{LinearProgramming}

\end{document}